\documentclass[12pt]{article}
\usepackage{epsf}
\def\ga{\mathrel{\mathpalette\fun >}}
\def\fun#1#2{\lower3.6pt\vbox{\baselineskip0pt\lineskip.9pt
\ialign{$\mathsurround=0pt#1\hfil
##\hfil$\crcr#2\crcr\sim\crcr}}}
\newcommand{\lan}{\left\langle}
\newcommand{\ran}{\right\rangle}

\newcommand{\ven}{\mbox{\boldmath${\rm n}$}}

\newcommand{\be}{\begin{equation}}
\newcommand{\ee}{\end{equation}}
\newcommand{\bc}{\begin{center}}
\newcommand{\ec}{\end{center}}

\newcommand{\mr}[1]{\mathrm{#1}}
\newcommand{\fr}[2]{\frac{#1}{#2}}
\newcommand{\lt}{\left}
\newcommand{\rt}{\right}

\newcommand{\lb}{\label}

\title{\bf Static potential in baryon}
\author{D.S. Kuzmenko}
\date{\it Institute of Theoretical and Experimental
Physics,\\ 117218, B.Cheremushkinskaya 25, Moscow, Russia}

\begin{document}
\maketitle

\begin{abstract}
The baryon static potential is calculated in the framework of 
field correlator method and is shown to match the recent lattice 
results. The effects of the nonzero value of the gluon 
correlation length are emphasized.
\end{abstract} 

1. The static potential is a key quantity in the calculations
of spectra and wave functions of baryons in the potential models 
\cite{1}, as well as in the QCD string approach \cite{2}.
On the other hand the static potential is generated by the gluon 
forces of QCD and therefore sheds the light on the profound 
properties of strong interactions.   Accurate numerical 
calculations of static potential in baryon performed in lattice 
QCD  last years \cite{3}, \cite{4} induced new significant growth 
of interest to the structure of the gluon forces in baryon. 

The results of calculations of the baryon potential in 
the framework of the method of the field correlators (MFC) 
\cite{5} are presented in this report. 
We set the correspondence of this
potential to lattice one and underline fundamental effects of
strong interactions --- the confinement of color charges and
definite value of the correlation length of the gluon fields,
which reveal themselves in the behavior of baryon potential.

2. The static potential is expressed through the Wilson loop
${\cal W}_B$ according to the relation 
\be
V_B=
-\lim_{T\to \infty} \frac{1}{T}\ln  {\cal W}_B,
\label{1}
\ee
and that is why we are proceeding now to the calculations
of Wilson loop.
The baryon Wilson loop is shown in Fig. 1 and is defined as  
follows,
\be
{\cal W}_B=\lan \frac16 \epsilon_{\alpha\beta\gamma}
\epsilon^{\alpha '\beta '\gamma '}
\Phi^{\alpha}_{\alpha '}(C_1)\Phi^{\beta}_{\beta '}(C_2)
\Phi^{\gamma}_{\gamma '}(C_3)\ran,
\label{2}
\ee
where the parallel transporter or Schwinger line $\Phi$ is given 
as
\be
\Phi^{\alpha}_{\beta}(x,y,C)= (P\exp ig\int_C A_\mu d
z_\mu)^{\alpha}_{\beta}.
\lb{3}
\ee
The average of the Wilson loop over the vacuum fields using
the nonabelian Stokes theorem and bilocal (two-point)
approximation is performed in  \cite{5}, and the result reads as
$$
{\cal W}_B=\exp\left\{-\sum_{i=1}^3\frac12\int_{S_i}\int_{S_i}
 d\sigma_{\mu_1\nu_1}(x) d\sigma_{\mu_2\nu_2}(x') 
{\cal D}_{\mu_1\nu_1,\mu_2\nu_2}(x-x')+\right. 
$$   
\be
\left.\sum_{i<j}\frac12\int_{S_i}\int_{S_j}
 d\sigma_{\mu_1\nu_1}(x) d\sigma_{\mu_2\nu_2}(x') 
{\cal D}_{\mu_1\nu_1,\mu_2\nu_2}(x-x') \right\},
\label{4}
\ee
where ${\cal D}$ designates the bilocal field strength 
correlators, 
\be
{\cal D}_{\mu_1\nu_1,\mu_2\nu_2}(x-x')\equiv
\frac{g^2}{N_c}\,\mathrm{tr} \langle F_{\mu_1\nu_1}(x)\Phi(x,x')
F_{\mu_2\nu_2}(x')   \Phi(x',x)\rangle,
\lb{5}
\ee
and it is assumed that the straight-line trajectories for 
parallel transporters are chosen.
Integrations in (\ref{4}) are taken over the surfaces $S_i$
of the Wilson loop, which are formed by the trajectories of the 
corresponding quark, $C_i$,  and that of the string 
junction, see Fig. 1. The trajectory of the string junction is 
shown in the figure by the dotted line.  

Relying on the phenomena of the Casimir scaling \cite{6}, which 
is confirmed in lattice simulations, one can expect that the 
bilocal approximation (\ref{4}) is valid within the accuracy of 
1\%. 

\begin{figure}[!t]
\epsfxsize=8cm
\hspace*{4.35cm}
\epsfbox{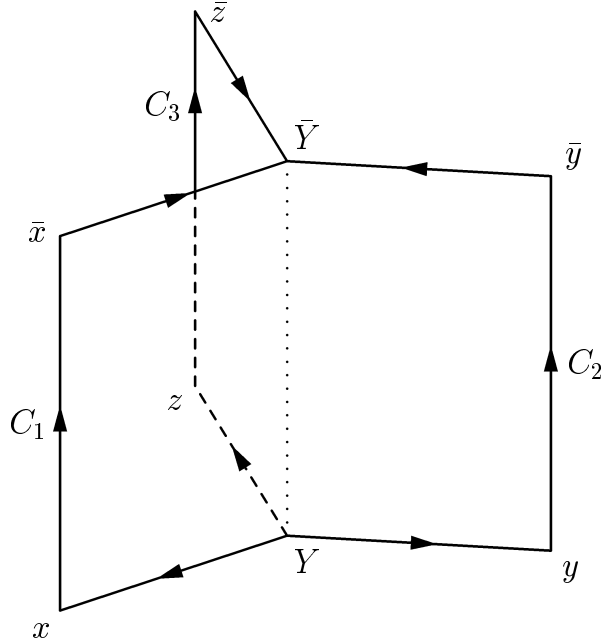}
\caption{The baryon Wilson loop}
\end{figure}

 Performing the surface integration in (\ref{4}) and 
substituting the result into (\ref{1}) we arrive at the
expression for the baryon potential in bilocal approximation,
\be 
V_B(R_1,R_2,R_3)=
\left(\sum_{a=b}-\sum_{a<b}\right)n_i^{(a)}n_j^{(b)}\int_0^{R_a}\int_0
^{R_b } d l\,d l' \int_0^\infty  d t\, {\cal D}_{i4,j4}(z_{ab}),
\label{6}
\ee
where $R_a$ is separation of the string junction and 
corresponding ($a$-th) quark, $\ven^{(a)}$ the unity vector
directed from the string junction to this quark, and
$z_{ab}=(l\,\ven^{(a)}-l'\ven^{(b)},t)$.

 According to MFC \cite{7}, the bilocal correlators are
written in the general form containing two scalar formfactors
$D(z)$ and $D_1(z)$,
$$
{\cal D}_{\mu_1\nu_1,\mu_2\nu_2}(z)
=(\delta_{\mu_1\mu_2}\delta_{\nu_1\nu_2}-
\delta_{\mu_1\nu_2}\delta_{\mu_2\nu_1})~D(z)~+ 
$$
\be
   \frac12\lt(\fr{\partial}{\partial z_{\mu_1}}
(z_{\mu_2}\delta_{\nu_1\nu_2}-z_{\nu_2}\delta_{\nu_1\mu_2})+
\fr{\partial}{\partial z_{\nu_1}}
(z_{\nu_2}\delta_{\mu_1\mu_2}-
z_{\mu_2}\delta_{\mu_1\nu_2})\rt)\,D_1(z). 
\lb{7}
\ee

The formfactor $D(z)$ decreases exponentially, 
\be
D(z)= D(0) \exp \lt(-\fr{|z|}{T_g}\rt),
\lb{8}
\ee
which reflects the stochastic properties of the nonperturbative 
background gluon fields and is justified by the lattice 
computations \cite{8}. This behavior leads to the asymptotic 
area law for the Wilson loop, where the string tension $\sigma$
is expressed through $D(z)$ as follows,
\be 
\sigma=\frac{\pi}{2}\int_0^{\infty} dz^2
D(z)=\pi D(0)\,T_g^2= 0.18\mbox{ GeV}^2.
\lb{9}
\ee
The string tension is the main nonperturbative parameter of MFC. 
Its value is defined phenomenologically by the slope of the meson  
Regge trajectories  \cite{9} and is directly related to the 
radius of confinement.

There is another parameter in (\ref{8}), the correlation length
of the background  gluon field $T_g$.
However, it is not an independent parameter. 
Its value is extracted from the energy of the gluon excitation of
the hadron spectra \cite{10} and may be calculated in the QCD 
string approach \cite{10}, \cite{11} using the only 
parameter $\sigma$. The energy of the gluon excitation
is large, $\sim 1.5$ GeV \cite{11}, and therefore the gluon 
correlation scale is significantly  less than the confinement 
scale. 

The formfactor $D_1(z)$ is dominated by the contribution of
one-gluon-exchange of the perturbation theory, which gives rise to 
the static color-Coulomb potential. In the case of the baryon with 
the quarks forming equilateral triangle with the side $r$ 
the perturbative potential reads as
\be
V_{\mr{pert}}(r)=-\frac32\,\frac{C_F\alpha_s}{r},
\lb{10}
\ee
where $C_F=4/3$.

\begin{figure}[!t]
\epsfxsize=12cm
\hspace*{2.35cm}
\epsfbox{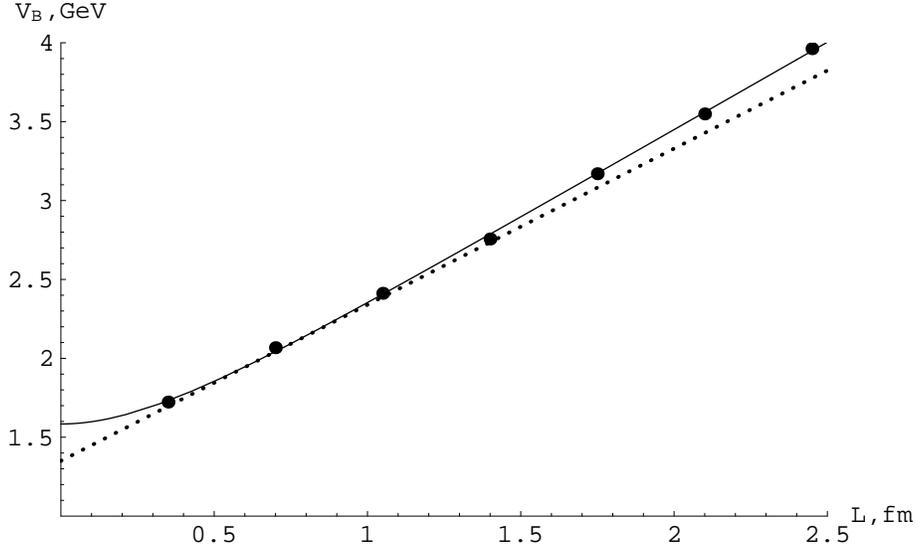}
\caption{The lattice nonperturbative baryon potential
from \cite{2} (points) for lattice parameter $\beta=5.8$ and  MFC
potential $V^{(B)}$ (solid line) with parameters   $\sigma=0.22$
GeV$^2$ and  $T_g=0.12$ fm vs. the minimal length of the string $L$.
The dotted line is a tangent at $L=0.7$ fm.}
\label{fig2}
\end{figure}

\begin{figure}[!t]
\epsfxsize=12cm
\hspace*{2.35cm}
\epsfbox{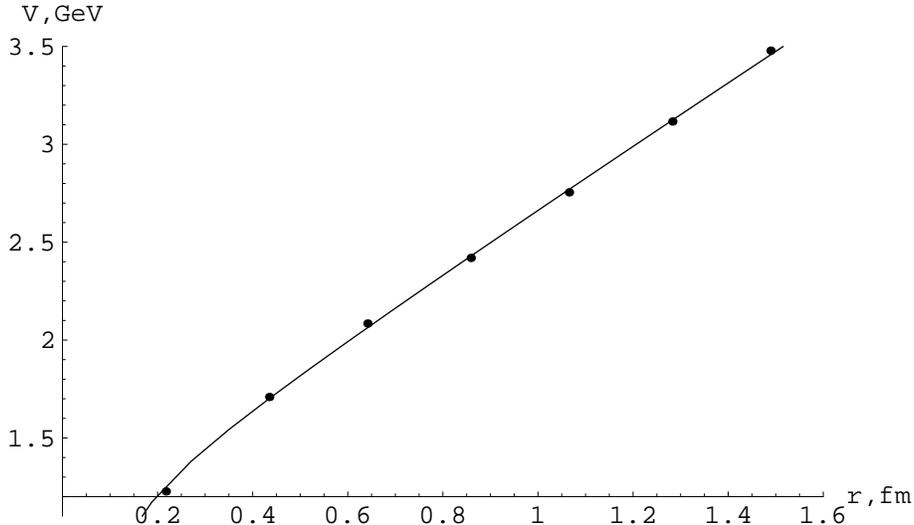}
\caption{The lattice baryon potential in the equilateral triangle with
quark separations $r$ from \cite{3} (points) at $\beta=5.8$
and the MFC potential $V^{(B)}+V^{\mathrm{pert}}_{\mathrm{(fund)}}$
(solid line) at  $\alpha_s=0.18$, $\sigma=0.18$ GeV$^2$, and
$T_g=0.12$ fm.}
\label{fig3}
\end{figure}

3. We now proceed to the comparison of the potential 
(\ref{6})-(\ref{10}) with the lattice results \cite{3}, \cite{4}.
 In Fig. 2 the results of the lattice 
simulations \cite{3} of the baryon potential with the perturbative 
part subtracted is shown by points in dependence on the 
length of the string, $L=R_1+R_2+R_3$.
Solid line in the figure shows the behavior of the nonperturbative 
potential calculated in MFC for the configurations of equilateral 
triangle.  
When $L\ga 1$ fm, the potential grows linearly having the slope 
$\sigma$. When the length of the string gets smaller,
the slope of the potential diminishes. Dotted line presents
a tangent to the MFC potential at $L=0.7$ fm. The slope of the 
tangent is $\sim 0.9\sigma$. Linear phenomenological potential 
with the same slope was used in the constituent model for the 
description of the spectra of baryons long time ago \cite{1}.
 One can see from the figure that  our curve goes
well through all lattice points. 
The configurations of quarks forming triangles with the angles in the region 
the region $[\pi/20,\pi/2]$, which were used in lattice work 
\cite{3}, do not allow to establish any dependence of the 
potential on the configuration with the accuracy given. 
The study of the MFC potential on quark configurations will be
performed below (see Fig. 4).
 
In Fig. 3 the results of lattice computations of the potential in 
equilateral triangle \cite{4} are presented (points)  vs. the 
quark separations $r$. The MFC potential with the perturbative 
part included is shown by solid line.  One can see
that all the lattice data are completely described by
the MFC potential. 

Computations in \cite{3}, \cite{4} are performed in quenched 
approximation. In the MFC calculations sea quarks are 
not considered too. The studies of  the effects of light dynamical 
quarks on the $Q\bar Q$ static potential in lattice were recently 
performed in \cite{12}. No clear evidence for a flattening of the 
potential was found up to distances 2.5 fm. One can therefore 
expect that the sea quarks would not change the baryon potential 
significantly at typical hadron sizes.

 It is also interesting to study the dependence of the potential 
in baryon  on the quark locations at fixed length of the string 
$L$. Let us consider isoceles  triangles with the full length 
of the string $L$ and vertex $\gamma$ and   denote $V^L(\gamma)$
the nonperturbative potential in these triangles.
At large enough sizes $L\ga 1$ fm$\gg T_g$ asymptotic relations
follow from (\ref{6})-(\ref{9}). When $\gamma=0$ and locations of 
two quarks coinside, these quarks  combine in antitriplet.  The 
string consists of one line and  
\be
V_1\equiv V^L(\gamma=0)= \sigma L-\frac4{\pi}\,\sigma T_g.
\label{11}
\ee
When $0<\gamma<2\pi/3$, the string consists of three lines,
and potential
\be
V_2\equiv V^L(0<\gamma<2\pi/3)= \sigma L+\left(-\frac{12}{\pi}
+\fr{2}{\sqrt{3}}\right)\sigma T_g.
\label{12}
\ee
When $\gamma\geq 2\pi/3$, the string consists of two lines.
In this case
\be
V^L(\gamma\geq 2\pi/3)= \sigma L+\left(-\frac8{\pi}
+\fr2{\pi}(\pi-\gamma)\,\mr{ctg}\,(\pi-\gamma)\right)\sigma T_g.
\label{13}
\ee

\begin{figure}[!t]
\epsfxsize=12cm
\hspace*{2.35cm}
\epsfbox{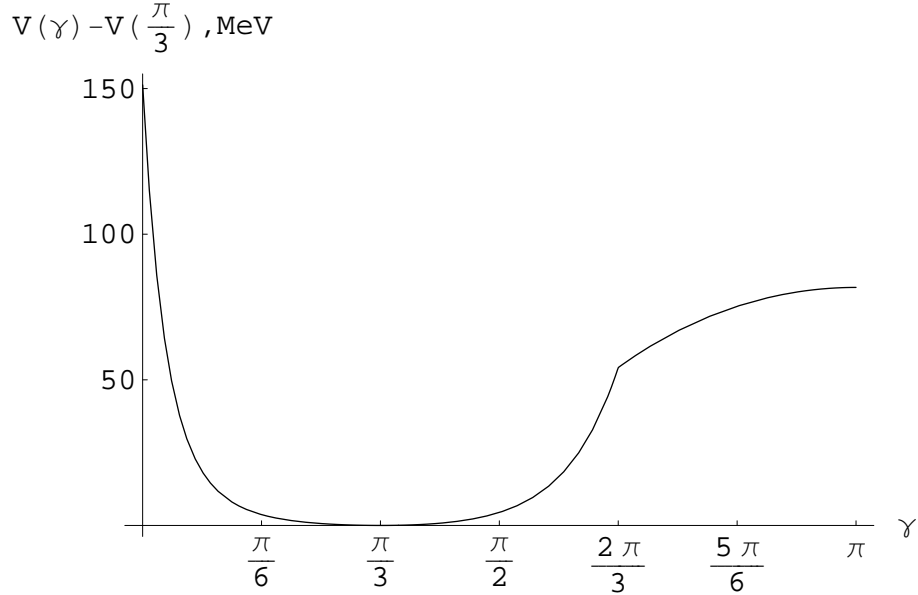}
\caption{The baryon potential in isoceles triangle with the length of the
string $L=1.8$ fm versus the vertex $\gamma$ for $\sigma=0.18$
GeV$^2$, $T_g=0.12$ fm.}
\label{fig4}
\end{figure}

It is not difficult to calculate difference between the 
potentials in different configurations (we use $\sigma$=0.18 
GeV$^2$,  $T_g$=0.12 fm),
\be
\Delta V_1\equiv V_1-V_2
=\left(\fr8{\pi}-\fr2{\sqrt{3}}\right)\sigma T_g
\approx 150 \mbox{ MeV},
\label{14}
\ee
\be
\Delta V_2\equiv V^L\lt(\fr{2\pi}3\right)-V_2
=\left(\fr4{\pi}-\fr{4}{3\sqrt{3}}\right)\sigma T_g
\approx 55 \mbox{ MeV},
\label{15}
\ee
\be
\Delta V_3\equiv V^L(\gamma\to\pi)-V^L\left(\fr{2\pi}3\right)=
\left(\fr2{\pi}-\fr{2}{3\sqrt{3}}\right)\sigma T_g
\approx 30 \mbox{ MeV}.
\label{16}
\ee
In Fig. 4 the dependence of the baryon potential in isoceles 
triangle on the angle $\gamma$ at $L=1.8$ fm is shown.
One can verify that for the curve shown in the 
figure  the relations (\ref{13})-(\ref{16}) are justified (taking 
$\gamma=\pi/3$ for $V_2$).

The nonperturbative potential in the configurations under 
consideration has a pike near $\gamma$=0. However, in this 
region  the perturbative part of the total potential 
dominates. Indeed, the effect reveals itself when
 the quark separation $r_{qq}$ becomes less than the background correlation 
length. But at  separation $r_{qq}\approx 0.1$ fm the  
 color-Coulomb interaction of quarks attains 300 MeV and grows 
rapidly when $r_{qq}$ diminishes.
That is why to answer the question about the relevance of 
quark-diquark configurations we are to consider the motion of two 
quarks in their perturbative potential. In particular, the radius 
of diquark may be estimated using the Bohr formula, 
$r_{qq}=2/(m_q C_F \alpha_s)$, which yields $r_{bb}\approx 0.3$ fm 
and $r_{cc}\approx 1.0$ fm. More accurate values were obtained
in the relativistic quark model \cite{13}, where relativistic
corrections to perturbative potential as well as linear 
nonperturbative potential were accounted for. The radii of 
diquarks calculated in \cite{13} for the ground states are
$r_{bb}= 0.37$ fm, $r_{cc}= 0.56$ fm. We are to state that
the radii of diquarks are comparable with the size of baryon, 
which perhaps signals that the formation of quark-diquark 
configuration is unprobable. Nevertheless, the quark-diquark 
approximation turns out usefull for the computations of the baryon 
spectra (see e.g. \cite{13}). Account of the 
nonperturbative effects considered in (\ref{11})-(\ref{16}) and 
Fig. 4 would spoil this approximation only a little.

4. To summarize, we have calculated nonperturbatively the static
potential in baryon and demonstrated that it completely describes 
recent lattice results.  The latter means in particular 
that the problem of the form of the potential is resolved.
The potential in our approach has apparent $Y$-type structure,
i.e. it depends only on the distances from the quarks to the 
string junction, the location of the latter being determined
by the condition of the minimal total length of the string. 
The formal answer of the problem of the potential law
is that the slope of the nonperturbative potential in dependence 
on the length of the string $L$ grows from zero to $\sigma$ when 
$L$ changes from zero to $\sim$1.5 fm. Therefore an effective 
slope at typical hadronic sizes may be chosen $\sim 0.9~\sigma$.

The behavior of the baryon potential considered is a 
consequence of two profound properties of the strong interaction, 
the confinement of quarks and the definite value of the 
correlation length of nonperturbative gluon fields. The latter is 
directly related to the energy scale of the gluon excitations of 
the hadron spectra.  It is the correlation length  that 
induces the change of the slope of the potential in baryon. 
Note that for mesons situation differs. The slope of 
 the nonperturbative static potential in mesons  almost does not 
change at small distances due to the interference with the 
perturbative fields \cite{14}.

Apart from the change of the slope of the potential in 
baryon, another related effect was studied, namely the behavior of 
the potential when the length of the string is fixed. 
It was demonstrated in particular that the difference of the 
potentials in configurations with the string consisting of 
different (one, two or three) numbers of lines turns out to be 
proportional to $\sigma T_g$. The combination of the parameters
directs immediately that the effect is induced by both the scale 
of confinement and correlations of gluon fields. 
The influence of the effect on the  creation 
of the quark-diquark configuration is shown to be small. 
Last but not least, it would be interesting to study this effect  
in devoted  lattice calculations. 

 The author is grateful to Yu.A.Simonov for many fruitful 
discussions. This work has been supported by 
 RFBR grants 00-02-17836,  00-15-96786, and INTAS 00-00110, 
00-00366.

\end{document}